\documentclass{emulateapj}
\usepackage{apjfonts}

\slugcomment{Received 2010 November 8; accepted 2011 February 7; published 2011 April 8}

\shorttitle{RADIO STRUCTURE OF PSR~B1259$-$63/LS~2883}
\shortauthors{Mold\'on et al.}

\begin{document}


\title{Discovery of extended and variable radio structure from the\\ gamma-ray binary system PSR~B1259$-$63/LS~2883}


\author{Javier Mold\'on\altaffilmark{1}, Simon Johnston\altaffilmark{2}, Marc Rib\'o\altaffilmark{1}, Josep M. Paredes\altaffilmark{1}, and Adam T. Deller\altaffilmark{3,4}}

\affil{$^{1}$Departament d'Astronomia i Meteorologia, Institut de Ci\`encies del Cosmos (ICC), Universitat de Barcelona (IEEC-UB), Mart\'{\i} i Franqu\`es 1, 08028 Barcelona, Spain
}
\affil{$^{2}$Australia Telescope National Facility, CSIRO, P.O. Box 76, Epping, NSW 1710, Australia
}
\affil{$^{3}$National Radio Astronomy Observatory (NRAO), P.O. Box 0, Socorro, NM 87801, USA
}
\affil{$^{4}$Astronomy Department, University of California at Berkeley, 601 Campbell Hall, Berkeley, CA 94720, USA
}




\begin{abstract}

\object{PSR~B1259$-$63} is a 48~ms pulsar in a highly eccentric 3.4 year orbit around the young massive star \object{LS~2883}. During the periastron passage the system displays transient non-thermal unpulsed emission from radio to very high energy gamma rays. It is one of the three galactic binary systems clearly detected at TeV energies, together with \object{LS~5039} and \object{LS~I~+61~303}. We observed \object{PSR~B1259$-$63} after the 2007 periastron passage with the Australian Long Baseline Array at 2.3~GHz to trace the milliarcsecond (mas) structure of the source at three different epochs. We have discovered extended and variable radio structure. The peak of the radio emission is detected outside the binary system near periastron, at projected distances of 10--20~mas (25--45~AU assuming a distance of 2.3~kpc). The total extent of the emission is $\sim$50~mas ($\sim$120~AU). This is the first observational evidence that non-accreting pulsars orbiting massive stars can produce variable extended radio emission at AU scales. Similar structures are also seen in \object{LS~5039} and \object{LS~I~+61~303}, in which the nature of the compact object is unknown. The discovery presented here for the young non-accreting pulsar \object{PSR~B1259$-$63} reinforces the link with these two sources and supports the presence of pulsars in these systems as well. A simple kinematical model considering only a spherical stellar wind can approximately trace the extended structures if the binary system orbit has a longitude of the ascending node of $\Omega\sim-40^{\circ}$ and a magnetization parameter of $\sigma\sim0.005$.

\end{abstract}



\section{Introduction}\label{introduction}

The binary system \object{PSR~B1259$-$63}/\object{LS~2883}\footnote{Star 2883 in the catalog of Luminous Stars in the Southern Milky Way \citep{stephenson71}. The use of SS~2883 should be avoided, see \cite{negueruela11}.} is formed by a young 48~ms radio pulsar in an eccentric orbit of 3.4~years around a massive main-sequence star \citep{johnston92, johnston94}. The parameters of the system are shown in Table~\ref{table:system}. The spectral type of the massive star, O9.5\,Ve, and some of the binary parameters have been recently updated by \cite{negueruela11}, who obtained a distance to the system of $2.3\pm0.4$~kpc. Close to the periastron passage the system displays non-thermal unpulsed emission that has been detected in radio \citep{johnston05}, X-rays \citep{cominsky94, uchiyama09, chernyakova09}, hard X-rays up to 200~keV \citep{grove95}, and very high energy (VHE; 0.1--100~TeV) $\gamma$-rays above $380$~GeV \citep{aharonian05, aharonian09}. In the range $\sim$0.1--100~GeV strict upper limits were obtained by EGRET close to the 1994 periastron passage \citep{tavani96}, and the source has not yet been observed by \textit{AGILE} and \textit{Fermi} close to periastron. The VHE emission is variable on orbital timescales, and is interpreted as the result of inverse Compton upscattering of stellar UV photons by relativistic electrons, which are accelerated in the shock between the relativistic wind of the young non-accreting pulsar and the wind of the stellar companion (see \citealt{maraschi81, tavani97, kirk99, dubus06, bogovalov08}, and references therein).

The high-mass binaries \object{LS~5039} and \object{LS~I~+61~303} have also been detected at VHE, and show a broadband spectral energy distribution (SED) similar to that of \object{PSR~B1259$-$63}/\object{LS~2883} \citep{dubus06}. In contrast to X-ray binaries, all three sources have the peak of the SED at MeV-GeV energies. For these reasons, they can be considered gamma-ray binaries. However, the nature of the compact objects in \object{LS~5039} and \object{LS~I~+61~303} is unknown because their masses are not well constrained by the system mass functions \citep{casares05a, casares05b, aragona09}, and no pulsations have been found. These systems have been extensively observed at VHE during several orbital cycles, while the observations of \object{PSR~B1259$-$63}, the only system with a confirmed pulsar, are scarce due to the long orbital period. Any observational link between the three gamma-ray binaries would shed light in the understanding of this kind of systems. 

\begin{deluxetable*}{l c c c } 
\tablecaption{Parameters of the Pulsar, the Massive Star and the Binary System. 
\label{table:system}}
\tablehead{
\colhead{Parameter} & \colhead{Symbol} & \colhead{Value} & \colhead{Reference}}
\startdata
Pulsar period           & $P$           & $47.762506780(2)$~ms                  & 1        \\
Period derivative       & $\dot{P}$     & $2.276554(2)\times 10^{-15}$          & 1        \\
Characteristic age      & $\tau_{c}$    & $3.3\times 10^5$~yr                   & 2        \\
Surface magnetic field  & $B$           & 3.3$\times 10^{11}$~G                 & 2        \\
Spindown luminosity     & $\dot{E}_{\rm sp}$& $8\times 10^{35}$~erg~s$^{-1}$    & 3        \\
Spectral type           & \nodata            & O9.5\,Ve                          & 4        \\
Effective temperature   & $T_{\rm eff}$ & $27\,500$--$34\,000$~K                & 4        \\
Surface gravity         & $\log~g$      & $3.7$--$4.1$                          & 4        \\
Radius                  & $R_{1}$       & $8.1$--$9.7~R_\odot$                  & 4        \\
Optical luminosity      & $L_{\rm opt}$ & $2.4\times10^{38}$~erg~s$^{-1}$       & 4        \\
Mass                    & $M_{1}$       & $31~M_\odot$                          & 4        \\
Distance                & $d$           & $2.3\pm0.4$~kpc                       & 4        \\
Mass function           & $f(M_{2})$    & 1.53~$M_\odot$                        & 5        \\
Terminal wind velocity  & $v_{\infty}$  & $1350\pm200$~km~s$^{-1}$              & 6        \\
Orbital period          & $P_{\rm orb}$ & 1236.72432(2)~days                    & 1        \\
Reference epoch         & $T_{0}$       & MJD~48124.34911(9)                    & 1        \\
Semimajor axis          & $a_{\rm 2}$   & $7.2\pm0.4$~AU                        & 4\tablenotemark{a}        \\
Orbit inclination             & $i$           & $22\fdg2{\pm1.4}$                     & 4\tablenotemark{a}        \\
Eccentricity            & $e$           & 0.8698872(9)                          & 1        \\
Argument of periastron  & $\omega_{\rm 2}$      & $138\fdg6659(1)$         & 1        \\
Longitude of ascending node & $\Omega$  & $-40^{\circ}$                         & See the text \\
Proper motion (rigth ascension) & $\mu_{\alpha}\cos\delta$ & $-1.4\pm2.7$~mas~yr$^{-1}$  & 7 \\
Proper motion (declination)     & $\mu_{\delta}$           & $-3.2\pm1.9$~mas~yr$^{-1}$  & 7 \\
\enddata
\tablecomments{The values in parentheses refer to the uncertainty in the last digit at 1$\sigma$ level.}
\tablenotetext{a}{Derived from (4).}
\tablerefs{(1) \citealt{wang04}; (2) \citealt{wex98}; (3) \citealt{manchester95}; (4) \citealt{negueruela11}; (5) \citealt{johnston94}; (6) \citealt{mccollum93}; (7) \citealt{zacharias09}.}
\end{deluxetable*}

The radio emission from the \object{PSR~B1259$-$63} system is described most recently in \cite{johnston05}, which includes multiwavelength data from ATCA observations obtained during the periastron passages of 1994, 1997, 2000, and 2004 (hereafter we will refer to the epoch of each periastron passage as $\tau$). The emission has two non-thermal components. The first one is pulsed emission with a flux density of $\sim$2--5~mJy at 2.5~GHz and a nearly flat spectral index, which disappears approximately from 16~days prior to periastron ($\tau-$16) to 15~days after periastron ($\tau+$15). This has been interpreted as an eclipse of the pulsar when it crosses behind the equatorial circumstellar disk present around the massive star \citep{melatos95}. The second component is transient unpulsed synchrotron emission that appears at $\tau-$20~days and shows two peaks centered around $\tau-$10 and $\tau+$20~days, with flux densities at 2.5~GHz up to $\sim$15--20~mJy and $\sim$30--50~mJy, respectively. After the post-periastron peak, the flux density of the unpulsed emission decreases continuously, and it has been detected up to $\sim\tau+$100~days. This transient emission remains optically thin during the outbursts.

The broadband transient emission of \object{PSR~B1259$-$63}/ \object{LS~2883} is produced around periastron, when strong interaction is produced between the stellar and pulsar winds. The shocked material is contained by the stellar wind behind the pulsar, producing a nebula extending away from the stellar companion. Along this adiabatically expanding flow, the accelerated particles produce synchrotron emission from radio to X-rays \citep{tavani97, kirk99, dubus06, takata09}. The expected morphology depends on the magnetization parameter of the pulsar wind, $\sigma$, defined as the upstream ratio of magnetic to kinetic energy. These models predict that the radio emission extends up to several AU (corresponding to several milliarcseconds, mas, at 2.3~kpc), and that its structure should be variable on orbital timescales. This radio behavior has not been tested in \object{PSR~B1259$-$63}, which only emits over a period of a few months every 3.4~year.

Here we present the first VLBI radio images of \object{PSR~B1259$-$63}, obtained close to its 2007 periastron passage. The high-resolution images at three different orbital phases provide a direct view of the small-scale morphology of the source, which is comparable to those previously observed in \object{LS~5039} and \object{LS~I~+61~303}.

\section{Observations and data reduction}\label{observations}

\begin{figure*}[] 
\resizebox{1.0\hsize}{!}{\includegraphics[angle=0]{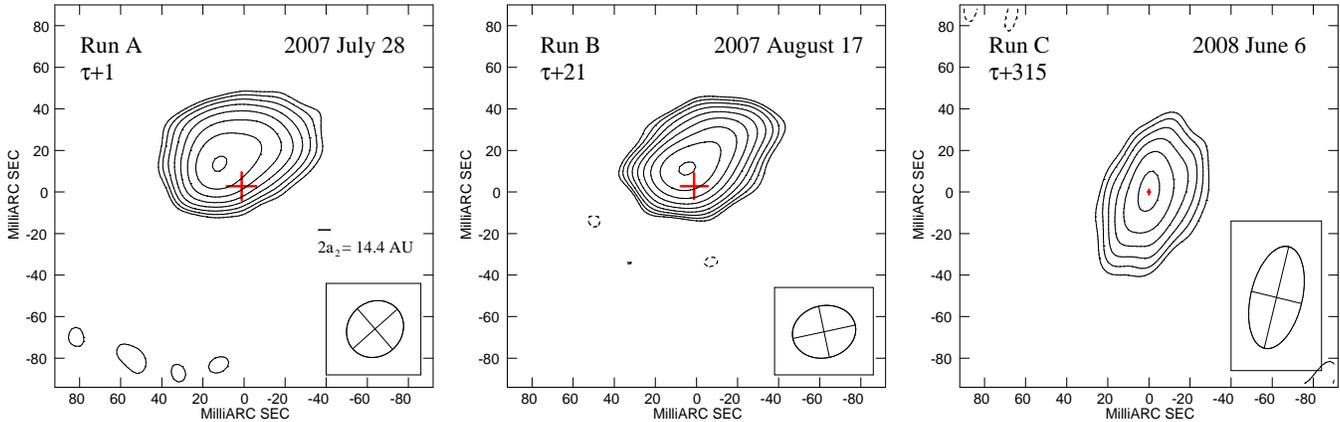}}
\caption{LBA images of PSR~B1259$-$63 at 2.3~GHz. North is up and east is to the left. The dates and the days after the periastron passage ($\tau$) are quoted at the top of each panel. The synthesized beam is displayed in the rectangle on the bottom-right corner of each image. The red crosses mark the region where the pulsar should be contained in each run (see the text). As a reference, the size of the major axis of the orbit of PSR~B1259$-$63/LS~2883 is shown in the first panel. For each image, the displayed contours start at 3$\sigma$ and increase by factors of $2^{1/2}$; the 1$\sigma$ rms close to the source in each image from left to right is 0.30, 0.66, and 0.15~mJy~beam$^{-1}$.
\label{fig:f1}}
\end{figure*}

\begin{deluxetable*}{c c c c c c c c } 
\tabletypesize{\scriptsize}
\tablecaption{Observational Parameters for Each Run. \label{table:observations}}
\tablewidth{0pt}
\tablehead{
\colhead{Run} & \colhead{MJD} & \colhead{On-source Time} & \colhead{No. of Antennas}   & \colhead{$\tau+$days\tablenotemark{a}} & \colhead{Orbital Phase\tablenotemark{a} ($^\circ$)} & \colhead{True Anomaly ($^\circ$)\tablenotemark{a}} & \colhead{$\theta$($^\circ$) \tablenotemark{a}}
}
\startdata
A  & 54309.04--54309.46  & 4.65 & 5 &   1.1(2)--1.5(2)   & 0.00087(14)--0.00120(14) &   9.1(1.4)--12.5(1.4) & 191 \\
B  & 54328.97--54329.40  & 3.75 & 5 &  21.0(2)--21.4(2)  & 0.01698(14)--0.01733(14) &  98.3(3)--99.1(3)     & 279 \\
C  & 54623.29--54623.67  & 4.20 & 4 & 315.3(2)--315.7(2) & 0.25496(14)--0.25527(14) & 165.97(1)--165.99(1)  & 346 \\
\enddata
\tablenotetext{a}{Periastron passage at $\tau=$ MJD 54307.9710(1), from the fourth orbital solution obtained in \cite{wang04}.}
\tablecomments{$\theta$ is the mean true anomaly plus 180$^{\circ}$.}
\end{deluxetable*}

\object{PSR~B1259$-$63} was observed with the Australian Long Baseline Array (LBA) at 2.3~GHz (13~cm) on three epochs: 2007 July 28 (run~A), 2007 August 17 (run~B), and 2008 June 6 (run~C). The LBA observations were performed with five~antennas of the array: Parkes, ATCA, Mopra, Hobart (not present in run~C) and Ceduna. The observational parameters of each of the $\sim10$ hr runs are shown in Table~\ref{table:observations}. The small number of antennas provides a rather poor $uv$-coverage that makes the data calibration and imaging more difficult than for arrays with more elements, because we have less baseline redundancy. The data were recorded at a bit rate of 512~Mbps per telescope distributed in eight sub-bands (four for each right- and left-handed polarization) with a bandwidth of 16~MHz, each of them correlated using 64~frequency channels, two-bit sampling, and 2~s of integration time. Hobart and Ceduna recorded at 256~Mbps (only two sub-bands for polarization). The data were correlated at Swinburne University using the DiFX software correlator \citep{deller07} without applying pulsar gating or binning.

The observations were performed using phase referencing on the calibrator J1337$-$6509 (B1334$-$649), which has an angular separation of $4\fdg0$ from \object{PSR~B1259$-$63} and was correlated at $\alpha_{\rm J2000.0}=13^{\rm h} 37^{\rm m} 52\fs4443$ and $\delta_{\rm J2000.0}=-65\degr 09\arcmin 24\farcs900$. This reference position from the global VLBI solution 2006d\_astro\footnote{http://lacerta.gsfc.nasa.gov/vlbi/solutions/2006d/2006d\_apr.src} has an uncertainty of 13~mas. The cycle time was 6~minutes, spending half of the time on the phase calibrator and the target source alternatively. The total flux density of the phase calibrator was $457\pm11$, $367\pm20$, and $491\pm6$~mJy for runs~A, B, and C, respectively. The source J0538$-$4405 (B0537$-$441) was used as a fringe finder for runs~A and B, and J1337$-$6509 (B1349$-$439) was used for run~C. No astrometric check source was observed during the runs.

The data reduction was performed in AIPS\footnote{The NRAO Astronomical Image Processing System. http://www.aips.nrao.edu/}. Total electron content models based on GPS data obtained from the CDDIS data archive\footnote{The Crustal Dynamics Data Information System http://cddis.nasa.gov/} were used to correct phase variations due to the ionosphere. Standard instrumental corrections were applied (parallactic angle, instrumental offsets, and slopes between and within bands and bandpasses). Fringe fitting on the phase calibrator was performed with the AIPS task FRING, and the solutions were applied to the target source. A self-calibrated model of the calibrator was used as an input model for CALIB, which was used to reduce the amplitude instabilities on timescales greater than 10~minutes. The data were averaged in frequency and time, and clean images were produced. For run B only, a single round of phase self-calibration was applied, to mitigate the residual atmospheric phase instabilities, which were more noticeable at this epoch. The final images were produced using a natural weighting scheme (robust 5 within the AIPS task IMAGR). For run~B, robust 2 and tapering were applied to avoid the presence of possible unreliable high-resolution features due to sidelobes of the synthesized beam. Self-calibration slightly affects measured properties such as extent and position angle (P.A.), while it preserves the morphology.

\section{Results}\label{results}

The resulting VLBI images at 2.3~GHz are shown in Figure~\ref{fig:f1}. Extended emission is detected at distances up to 50--55~mas ($120$--$130\pm20$~AU at $2.3\pm0.4$~kpc) during the two runs shortly after the periastron passage. The emission becomes gradually fainter from the peak toward the northwest, and no individual components have been found. The P.A. of the extended emission with respect to the peak is $\sim-67^{\circ}$ for run~A and $\sim-50^{\circ}$ for run~B. The emission in run~C, 315~days after the periastron passage, is dominated by a point-like source of a few mJy. 

\begin{figure*}[] 
\begin{center}
\epsfxsize=14cm
\epsffile{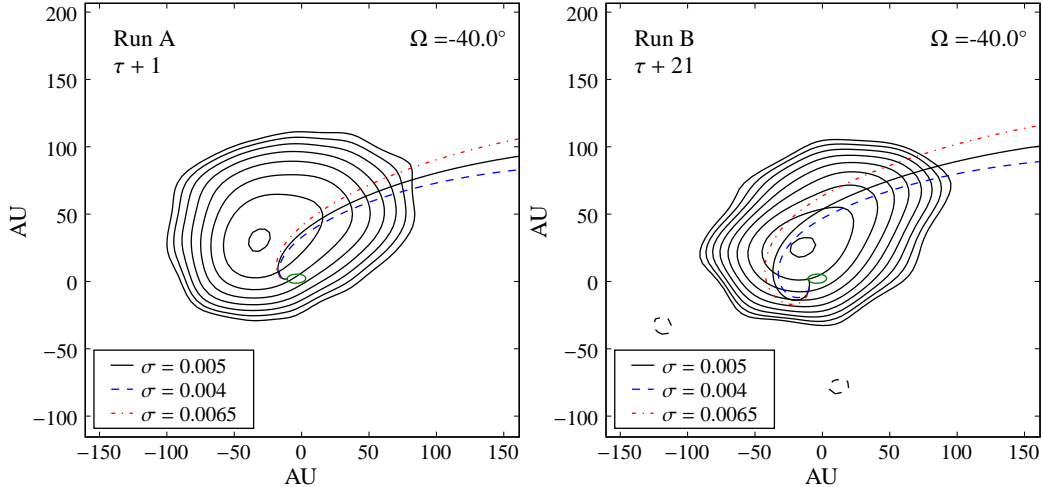}
\caption{Computed trajectory of the nebular flow in the past. The contour plots are the same as in Figure~\ref{fig:f1} (left and center). The green ellipse represents the counterclockwise orbit. The longitude of the ascending node, $\Omega$, is set to $-40^{\circ}$. The different magnetizations used are displayed in the bottom-left panels. The axes units are AU ($100$~AU $\simeq43$~mas).}
\label{fig:f2}
\end{center}
\end{figure*}

\begin{deluxetable*}{cccccccccccccc} 
\tabletypesize{\scriptsize}
\tablecaption{Source Parameters for the Images Shown in Figure~\ref{fig:f1}. 
\label{table:results}}
\tablewidth{0pt}
\tablehead{
\colhead{Run} & \colhead{$S_{\rm total}$} &\colhead{$S_{\rm peak}$} & \colhead{$\theta_{\rm HPBW}$}& \colhead{P.A.$_{\rm HPBW}$} & \multicolumn{2}{c}{Size} & \colhead{P.A.} & \colhead{$\Delta\alpha$} & \colhead{$\Delta\delta$} & \multicolumn{2}{c}{Separation} \\
\cline{6-7} \cline{11-12} \\
\colhead{} & \colhead{(mJy)} &\colhead{(mJy~beam$^{-1}$)} & \colhead{(mas)}& \colhead{($^{\circ}$)} & \colhead{(mas)} & \colhead{(AU)} & \colhead{($^{\circ}$)} & \colhead{(mas)}  & \colhead{(mas)} & \colhead{(mas)}& \colhead{(AU)}
}
\startdata
A  & 19.9 $\pm$ 1.4 & 10.4 $\pm$ 0.2 & 28.9 $\times$ 26.1 & $-49$ & 50  & 120 $\pm$ 20 & $-$67        & 11.3 $\pm$ 0.4 & 14.0 $\pm$ 0.5 & (10--20) $\pm$ 3 & (24--46) $\pm$ 7 \\
B  & 46.7 $\pm$ 1.0 & 32.7 $\pm$ 0.4 & 31.0 $\times$ 25.1 & $-78$ & 55  & 132 $\pm$ 22 & $-$50        &  4.2 $\pm$ 0.1 & 11.3 $\pm$ 0.1 & (5--14)  $\pm$ 3 & (12--31) $\pm$ 7 \\
C  & 3.0  $\pm$ 0.4 & 2.8  $\pm$ 0.4 & 50.3 $\times$ 25.1 & $-14$ & $<$2.8 & $<$6.7 $\pm$ 1.1 &\nodata&  0.0 $\pm$ 0.6 & 0.0  $\pm$ 1.1 & \nodata & \nodata \\
\enddata
\tablecomments{The columns are run label, total and peak flux density at 2.3~GHz, synthesized beam (HPBW) parameters, size and P.A. of the extended emission, the position offset of the peak of the emission from the reference position (position of run~C), and the range of possible separations between the peak and the pulsar position.}
\end{deluxetable*}

Like all VLBI arrays, the absolute flux calibration of the LBA relies on noise calibration injection and so the absolute flux values reported in Table~\ref{table:results} should be taken as uncertain at the $\sim10\%$ level. The flux densities of runs~A and B are compatible with previous ATCA observations at the corresponding orbital phases (the ATCA data from the current observations were not correlated as an independent array). The flux density in run~C is compatible with the flux density of the pulsar. This is expected considering the lack of unpulsed emission at $\tau+150$ and $\tau+180$ in previous ATCA observations \citep{johnston05}.

The phase-referenced observations allow us to obtain relative astrometry between runs. Since no astrometric check source was observed, we do not have real measurements of the astrometric uncertainties, and we will use the formal errors of a Gaussian fit obtained with JMFIT within AIPS. Given the relatively large calibrator--target separation of $\sim4^{\circ}$, we expect an additional systematic component to the position error due to the unmodeled ionosphere of 1--5~mas \citep{brisken00}. As a reference position for the plots we use the peak position of run~C, $\alpha_{\rm J2000.0}=13^{\rm h} 02^{\rm m} 47\fs6435(1)$ and $\delta_{\rm J2000.0}=-63\degr 50\arcmin 08\farcs636(1)$, which we consider to represent the pulsar position at MJD~54623.48. Assuming a mass of the neutron star of 1.4~$M_\odot$ and a stellar mass of 31$\pm5$~$M_\odot$ (see Table~\ref{table:system} for the system parameters), the mass function provides $i=22\fdg2$ and the semimajor axis of the pulsar orbit is $7.2$~AU, or $3.1$~mas for a distance to the system of $2.3$~kpc. The red crosses in Figure~\ref{fig:f1} mark the region where the pulsar should be located in each epoch. Their centers are placed at the position of run~C corrected for proper motion at the corresponding epochs (MJD 54309.25 for run A and MJD 54329.18 for run B). Since we do not know the orientation of the orbit in the sky, we plot as error bars the projected orbital separation of the pulsar with respect to run~C. The error bars also include the 1$\sigma$ uncertainties on the mass of the star (and hence the distance uncertainty), on the astrometry of run~C, and on the offset due to the proper motion. Finally, we have included the astrometry in runs~A and B to compute the range of possible separations between the peak of the emission and the pulsar, as shown in Table~\ref{table:results}. We consider all possible values of the longitude of the ascending node ($\Omega$), which determines the orientation of the orbit in the plane of the sky.

\section{Kinematical interpretation}\label{model}

The radio morphology at a given epoch depends on the spatial distribution of synchrotron emitting particles and their emission processes. Given the limitations of our data (only two images, and without accurate astrometry), we have used a simple kinematical model to check if it can trace the extended structures detected.

We have considered the shock between the relativistic pulsar wind and a spherical stellar wind. The shock is produced at the standoff distance, the region where the pulsar and stellar wind pressures balance, as described in \cite{dubus06}. The evolution of the nebular flow after the shock is described in \cite{kennel84}. This should be considered as a first approximation to the much more complex hydrodynamic behavior of a shocked flow in a binary system, as shown in \cite{bogovalov08}. In the Kennel \& Coroniti approximation of a non-turbulent adiabatically expanding flow, the flow speed depends only on the magnetization parameter $\sigma$ when assuming $\sigma\ll1$. This allows us to compute the past trajectory of the flow produced behind the standoff distance, which depends on the components separation along the orbit, the mass loss rate of the star ($0.6\times10^{-7}~M_\odot$~yr$^{-1}$ for a typical O9 star; \citealt{vink00}), the terminal wind velocity, $v_{\infty}$, and the spin-down luminosity of the pulsar, $\dot{E}_{\rm sp}$. With these restrictions, the only free parameters are the longitude of the ascending node, $\Omega$, which describes the orientation of the orbit, and the magnetization parameter, $\sigma$. Projecting the past trajectory of the flow on our VLBI images of runs~A and B, we found that the best match with the detected morphologies is obtained for $\Omega\simeq-40^{\circ}$ and a magnetization parameter of $\sigma\simeq0.005$, assuming an orbital inclination of $i=22\fdg2$. The obtained trajectories are shown in Figure~\ref{fig:f2} for three different values of $\sigma$. We note that the uncertainty in the distance to the system scales the size of the contours, but not the orbit and the trajectories, which are computed in AU. The range of magnetizations in the plots has been chosen to approximately show the effect of keeping $\sigma=0.005$ and changing either the distance from 1.9 to 2.7~kpc, either $\Omega$ from $-$35 to $-45^{\circ}$, or a variation of the product $\dot{M}~v_{\infty}$ of two orders of magnitude.

The parameters of the circumstellar equatorial disk of the star are uncertain, but considering $\dot{M}_{e}=5\times10^{-8}~\dot{M}$~yr$^{-1}$, \citep{johnston96}, and a wind velocity of $\sim$10~km~s$^{-1}$, the product $\dot{M}_{\rm e}~v_{\infty, \rm e}$ is $\sim$100 times smaller than the spherical wind contribution. If this equatorial component dominates during certain orbital phases, the flow trajectory would be closer to the orbit than the dashed models in Figure~\ref{fig:f2} for some specific regions of the more recent part of the trajectory. However, we do not attempt to accurately model this circumstellar disk, as its density, velocity, and crossing time are unknown. 

We note that the astrometric errors can be of the order of $\sim$10~AU (see above). We also emphasize that this simple model cannot account for the complex magnetohydrodynamical turbulences of the real flow and should be considered as a first approximation to constrain $\sigma$. Previous studies have assumed a magnetization parameter $\sigma$ of around 0.01--0.02 for this pulsar (see \citealt{tavani97}; \citealt{dubus06}), considerably higher than our best-fit value, although comparable lower values for $\sigma$ have been suggested for other systems \citep[e.g., the Crab pulsar;][]{kennel84}.

\section{Discussion and conclusions}\label{discussion}

The results presented in this Letter show that the particle accelerator within \object{PSR~B1259$-$63}/\object{LS~2883} can produce a flow of particles emitting synchrotron radiation that can travel several AU. The total projected extent of the nebula is $\sim50$~mas, or $120\pm20$~AU, and the peak of the emission is clearly displaced from the binary system orbit (see Figure~\ref{fig:f1} and Table~\ref{table:results}). Similar morphologies and displacements have been found in the other two known gamma-ray binaries, \object{LS~5039} and \object{LS~I~+61~303}, although for smaller sizes and on shorter timescales \citep{ribo08,moldon10,massi04,dhawan06,albert08}.

There is an ongoing debate on the nature of the compact object and particle acceleration mechanisms in \object{LS~5039} and \object{LS~I~+61~303}. Although initially suggested to be accreting/ejecting microquasar systems \citep{paredes00,massi04}, they are now thought to contain young non-accreting pulsars \citep{maraschi81,dubus06,dhawan06,ribo08,moldon08}, which can explain their multiwavelength emission \citep{sierpowska08, cerutti08, bogovalov08, zdziarski10}. However, there are three basic issues that are not well understood for \object{LS~5039} and \object{LS~I~+61~303}, and comparisons with \object{PSR~B1259$-$63}/\object{LS~2883} can help to clarify the situation. First, the putative pulsar properties of these two sources are unknown, as no pulsations have been detected for these systems. The lack of pulsations can be explained by the intense stellar wind that produces an extremely high absorption for these small orbits, a 3.9~day orbit with separations between 0.1 and 0.2~AU for \object{LS~5039}, and 26.5~days with separations of 0.1--0.7~AU for \object{LS~I~+61~303} \citep{casares05a,casares05b,aragona09}. As a reference, the separation for \object{PSR~B1259$-$63} is in the range 0.9--13.4~AU, and the pulsations disappear for distances below $\sim1.6$~AU. Second, it is not clear if the massive stellar wind can confine the pulsar wind \citep{romero07}. VLBI images, like the ones presented here, can shed light on the shock conditions and geometry. Third, the observed SED and variability at GeV energies is not well understood \citep{abdo09_ls5039,abdo09_lsi,torres10}. \textit{Fermi} and \textit{AGILE} are observing \object{PSR~B1259$-$63} for the first time during the 2010 periastron passage, and are providing GeV data that can be compared with those already obtained for \object{LS~5039} and \object{LS~I~+61~303}. In this context, the high-resolution VLBI radio observations presented here establish a common link to test the similarities between the three systems.

In conclusion, our results provide the first observational evidence that non-accreting pulsars orbiting massive stars can produce variable extended radio emission at AU scales. Similar structures are also seen in \object{LS~5039} and \object{LS~I~+61~303}, in which the nature of the compact object is unknown because the detection of pulsations is challenging. The discovery presented here for the young non-accreting pulsar \object{PSR~B1259$-$63} reinforces the link with these two sources and supports the presence of pulsars in these systems as well. Planned LBA observations of \object{PSR~B1259$-$63} during the 2010 periastron passage will allow us to provide a full comparison with the behavior observed in these sources, which have been extensively monitored during several orbital cycles. We have also shown that the orientation of the orbit and the magnetization of the pulsar can be inferred from VLBI observations of the source. Several images at different orbital phases covering a wider range of true anomalies will allow for a complete modeling of the orbital changes of the extended emission. Finally, accurate VLBI observations of the pulsed emission during several orbits can provide the pulsar trajectory, from which we can directly obtain the proper motion of the binary system, the inclination and $\Omega$ of the orbit, and the distance to the system. 

\acknowledgments

J.M., M.R., and J.M.P. acknowledge support by DGI of the Spanish Ministerio de Ciencia e Innovaci\'on (MICINN) under grants AYA2010-21782-C03-01 and FPA2010-22056-C06-02.
This work has been supported by the Consejer\'{\i}a de Innovaci\'on, Ciencia y Empresa of Junta de Andaluc\'{\i}a as research group FQM-322, and excellence grant FQM-5418.
J.M. acknowledges support by MICINN under grant BES-2008-004564.
M.R. acknowledges financial support from MICINN and European Social Funds through a \emph{Ram\'on y Cajal} fellowship.
The Australian Long Baseline Array is part of the Australia Telescope which is funded by the Commonwealth of Australia for operation as a National Facility by CSIRO. A.T.D. is a Jansky Fellow of the National Radio Astronomy Observatory (NRAO). The NRAO is a facility of the National Science Foundation operated under cooperative agreement by Associated Universities, Inc.


{}

\end{document}